\documentclass[
twocolumn,
superscriptaddress,
showpacs,preprintnumbers,
 amsmath,amssymb,
 aps,
prapplied,
longbibliography
]{revtex4-2}

\usepackage{graphicx}% Include figure files
\usepackage{dcolumn}% Align table columns on decimal point
\usepackage{bm}% bold math
\usepackage[update,prepend]{epstopdf}
\usepackage{xcolor}
\usepackage{ulem}
\usepackage{hyperref}% add hypertext capabilities
\begin{document}

%\preprint{Arxiv version}

\title{Ridge polariton laser: different from a semiconductor edge-emitting laser}

\author{H.~Souissi}
\email{Hassen.Souissi@umontpellier.fr}
\affiliation{Laboratoire Charles Coulomb (L2C), Universit\'e de Montpellier, CNRS, Montpellier, France}

 \author{M.~Gromovyi}
 \affiliation{Centre de Nanosciences et de Nanotechnologies, CNRS, Universit\'e Paris-Saclay, France}
 \affiliation{UCA, CRHEA-CNRS, Rue Bernard Gregory, 06560 Valbonne, France}
 \author{T.~Gueye}
\affiliation{Laboratoire Charles Coulomb (L2C), Universit\'e de Montpellier, CNRS, Montpellier, France}
\author{C.~Brimont}
\affiliation{Laboratoire Charles Coulomb (L2C), Universit\'e de Montpellier, CNRS, Montpellier, France}
\author{L.~Doyennette}
\affiliation{Laboratoire Charles Coulomb (L2C), Universit\'e de Montpellier, CNRS, Montpellier, France}
 \author{D.D~Solnyshkov}
\affiliation{Universit\'e Clermont Auvergne, CNRS, Institut Pascal, Clermont-Ferrand, France}
 \affiliation{Institut Universitaire de France (IUF), 75231 Paris, France}
 \author{G.~Malpuech}
 \affiliation{Universit\'e Clermont Auvergne, CNRS, Institut Pascal, Clermont-Ferrand, France}
 \author{E.~Cambril}
 \affiliation{Centre de Nanosciences et de Nanotechnologies, CNRS, Universit\'e Paris-Saclay, France}
 \author{S.~Bouchoule}
 \affiliation{Centre de Nanosciences et de Nanotechnologies, CNRS, Universit\'e Paris-Saclay, France}
 \author{B. Alloing}
 \affiliation{UCA, CRHEA-CNRS, Rue Bernard Gregory, 06560 Valbonne, France}
 \author{S.~Rennesson}
 \affiliation{UCA, CRHEA-CNRS, Rue Bernard Gregory, 06560 Valbonne, France}
 \author{F.~Semond}
 \affiliation{UCA, CRHEA-CNRS, Rue Bernard Gregory, 06560 Valbonne, France}
 \author{J.~Z\'u\~niga-P\'erez}
\affiliation{UCA, CRHEA-CNRS, Rue Bernard Gregory, 06560 Valbonne, France}
 \author{T.~Guillet}
 \email{Thierry.Guillet@umontpellier.fr}
 \affiliation{Laboratoire Charles Coulomb (L2C), Universit\'e de Montpellier, CNRS, Montpellier, France}

\date{\today}

\begin{abstract}
%Maximum 600 characters
We experimentally demonstrate the difference between a ridge polariton laser and a conventional edge-emitting ridge laser operating under electron-hole population inversion. The horizontal laser cavities are $20-60\ \mu m$ long GaN etched ridge structures with vertical Bragg reflectors. We investigate the laser threshold under optical pumping and assess quantitatively the effect of a varying optically-pumped length. The laser effect is achieved for an exciton reservoir length of just $15\%$ of the cavity length, which would not be possible in a conventional ridge laser, with an inversion-less polaritonic gain about 10 times larger than in equivalent GaN lasers. This combination of a very short injection section and a strong gain paves the way to compact microlasers with nonlinear functionalities for integrated photonics.
%The modelling of the cavity free spectral range demonstrates the polaritonic nature of the modes.
% sub-millimeter
\end{abstract}

%\pacs{}% PACS, the Physics and Astronomy
%\keywords{}%Use showkeys class option if keyword
\maketitle

% Maximum 3750 words
\section{Introduction} 
There are two conditions required for a laser to operate. The first is population inversion for a given electronic transition, so that the emission of the electronic system at the lasing frequency overcomes absorption, which can eventually vanish if the electronic system is fully inverted. In any case, the gain of the fully inverted system cannot exceed the maximal absorption, as prescribed by the well-known Einstein relations relating the rates of absorption and stimulated emission \cite{McCumber_Einstein_1964,Roosbroeck_Photon_1954}.
The second condition for lasing to occur is that gain overcomes the losses, essentially the decay rate of the photonic cavity constituting the laser. 

Polariton lasers \cite{imamoglu_nonequilibrium_1996,A.Kavokin2003,kavokin_microcavities_2017}
are coherent emitters based on hybrid quasi-particles arising from the strong coupling between excitons and photons, so-called exciton-polaritons.
They are by construction close to the so-called exciton lasers introduced in the seventies to explain laser-like emission in certain semiconductors such as ZnO \cite{Haug_Exciton_1967}.
Their putative advantage over conventional lasers is that they do not require population inversion, but only that the net spontaneous scattering rate from an excitonic reservoir overcomes the decay rate of the polariton mode. Following the proposal of a polariton laser without population inversion \cite{imamoglu_nonequilibrium_1996}, the demonstration of polariton lasing \cite{dang_stimulation_1998} was initially described and discussed in terms of cavity losses and an effective polaritonic gain.
The focus was later brought to a thermodynamic description in terms of polariton condensation in the ground state of a cavity polariton system \cite{kasprzak_bose-einstein_2006,kasprzak_formation_2008} and, ultimately, to the competition between kinetic and thermodynamic aspects of the condensation process \cite{nelsen_lasing_2009, kammann_crossover_2012, li_excitonic_2013, jamadi_polariton_2016, sun_bose-einstein_2017}. One should notice that photon condensation can also be realized in large area surface emitting lasers \cite{Barland_Photon_2021}, or by relaxing the population inversion condition in a molecular media with large Stokes shift between emission and absorption \cite{klaers_bose-einstein_2010}.
Back to polariton lasers, the most striking evidence of an operation regime without population inversion is the observation of two successive lasing thresholds: polariton lasing and lasing with population inversion \cite{bajoni_polariton_2008, bajoni_polariton_2012}  \cite{tempel_characterization_2012,Pieczarka_Crossover_2022}.

The polariton waveguide geometry has been explored for about a decade \cite{walker_exciton_2013,Solnyshkov2014,ciers_propagating_2017,walker_ultra-low-power_2015,jamadi_edge-emitting_2018, Suarez-Forero_Electrically_2020,jamadi_competition_2019, Walker_Spatiotemporal_2019, DiPaola2021,Ciers2020}. In the context of the current work, it presents the advantage to allow a selective excitation of only a part of the horizontal optical cavity confining the laser mode and revisit, thereby, laser interpretative frameworks, such as the balance between gain in the excited section and absorption.

In this work we study GaN-based deep-ridge waveguides delimited by GaN/air distributed Bragg reflectors (DBRs) and supporting polariton modes. The laser devices display a design very similar to the most common ridge semiconductor lasers, including recent InGaN-based devices \cite{Zhang_Short_2019}, however with much shorter cavities. We observe that lasing occurs while pumping only $15\%$ of the cavity length, {\it i.e.} much less than half of a cavity filled with an homogeneous active medium. This demonstrates the absence of reciprocity between gain and absorption in our system and represents an alternative and very clear demonstration of the occurrence of polariton lasing. An important outcome of injection sections displaying much stronger gain is the potential reduction of the devices footprint, which is of paramount importance in photonic integration. Furthermore, these polaritonic waveguides benefit from photonic nonlinearities orders of magnitude larger than in conventional nonlinear media due to polaritons dipolar interactions \cite{ Walker_Spatiotemporal_2019, DiPaola2021,Ciers2020}, opening the door for cointegration of laser sources and nonlinear functionalities (e.g modulation).

\section{Sample and experiment}

\subsection{Sample design}
\begin{figure}[h]
\resizebox{\hsize}{!}{\includegraphics{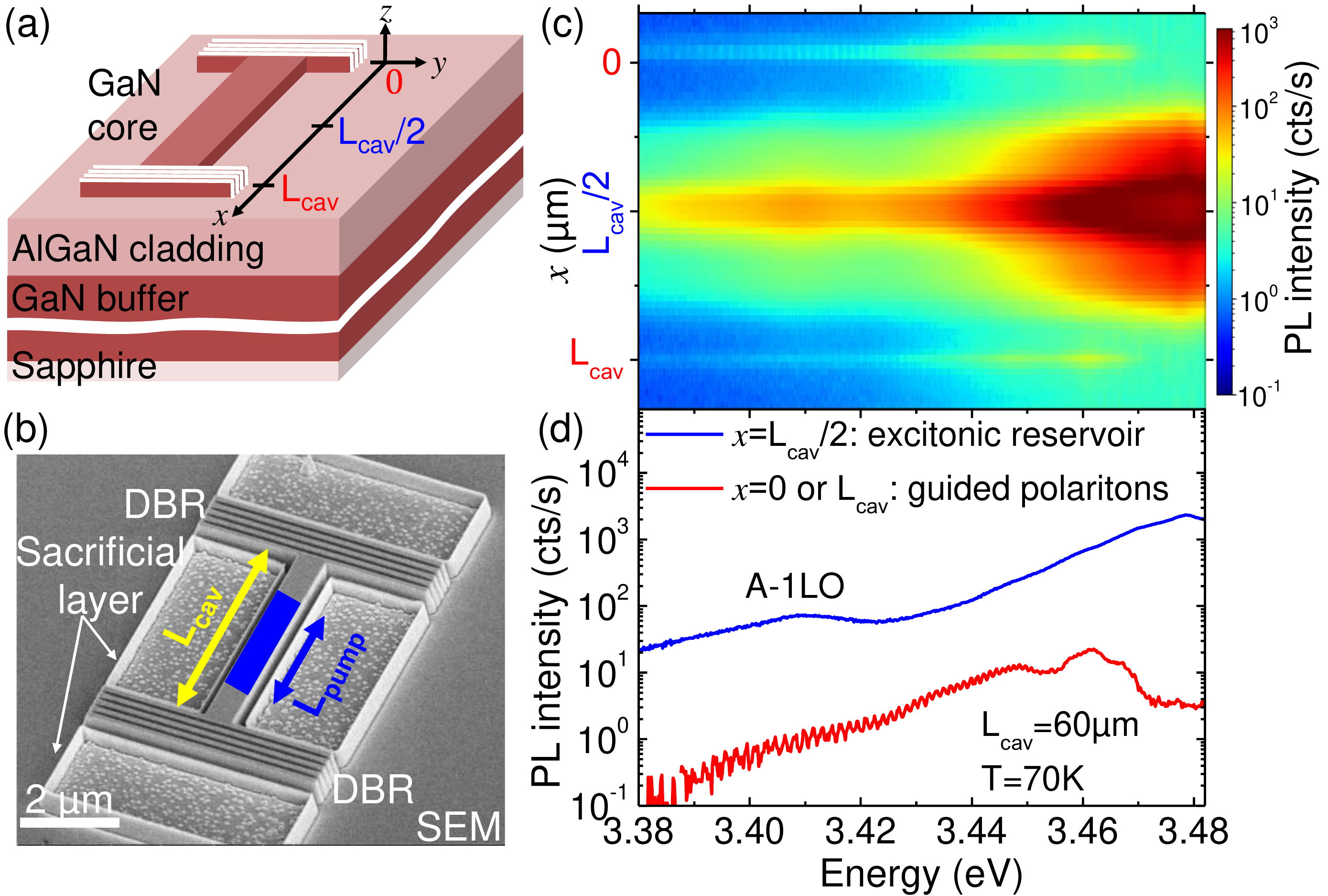}} 
  \caption{Sample description. (a)~Sample structure: ridge cavity after lithography and deep etching, the guided mode propagates along the $x$-axis in the GaN core; (b)~Scanning electron micrograph image of a $5\mu m-$long polariton ridge waveguide. The adjustable pump spot is sketched in blue; (c)~$\mu PL$ image of a $60\ \mu m$-long cavity excited below lasing threshold, with a line-shaped excitation spot ($L_{pump}=6.5~\mu m$); (d)~Emission spectra extracted from (c) at positions $x=L_{cav}/2$ (excitation spot) and $ x=0$ or $L_{cav}$ (DBR).}
  \label{fig:sample description}
\end{figure}

The sample consists in a GaN waveguide (WG) grown by metal-organic vapor phase epitaxy (MOVPE) on a c-plane sapphire. The structure is detailed in Figure~\ref{fig:sample description}.(a), with a 3-$\mu m$-thick GaN buffer, a 1.5-$\mu m$-thick Al$_{0.08}$Ga$_{0.92}$N cladding and a 150-nm-thick GaN waveguide core. The waveguide is almost monomode regarding the vertical confinement, with a confinement factor of $76\%$ for the zeroth-order transverse electric mode (TE0). The planar slab waveguide is then patterned by electron beam lithography, a dry deep etching and a complementary focused ion beam etching, so that the ridge waveguide with GaN/air DBRs benefits simultaneously from lateral optical confinement and efficient mirrors (Fig.~\ref{fig:sample description}(a,b)). In this work, we will focus on two devices with a $1\ \mu m$ ridge width, 4-pair DBRs, and a cavity length $L_{cav} = 20\ \mu m$ and $60\ \mu m$.

\subsection{Cavity imaging and spectroscopy}
The optical investigations are performed by microphotoluminescence imaging (µPL, see Appendix~\ref{app:experiment}). The cavity is pumped in its middle with a pulsed laser $(355\ nm,$ $7\ kHz$ rate, $4\ ns$ pulses) resonantly exciting the exciton reservoir with a line-shaped spot profile (Fig.~\ref{fig:sample description}(b)) which length $L_{pump}$ can be adjusted to excite partially or entirely the cavity, therefore controlling the size of the exciton reservoir. Figure~\ref{fig:sample description}(c) shows a spatially- and spectrally-resolved µPL image of the $60\ \mu m$-long cavity collected perpendicular to the sample, and for a $6.5\ \mu m$-long pump. At the position of the excitation spot ($x=L_{cav}/2$), the spectrum consists in a broad emission associated to the excitonic reservoir and its optical phonon replicas (Fig.~\ref{fig:sample description}(d), blue spectrum). At the position of the DBRs $ x=0$ or $L_{cav}$) acting as reflectors but also as scatterers towards the microscope objective, the specific emission is attributed to the guided polaritons (Fig.~\ref{fig:sample description}(d), red spectrum)
, with a series of sharp peaks corresponding to the Fabry-Perot (FP) modes of the cavity. Their free spectral range (FSR) is inversely proportional to the cavity length, as shown in Appendix~\ref{app:laser60mum}.

\section{Results}
\subsection{Laser operation}
\begin{figure}
\resizebox{\hsize}{!}{\includegraphics{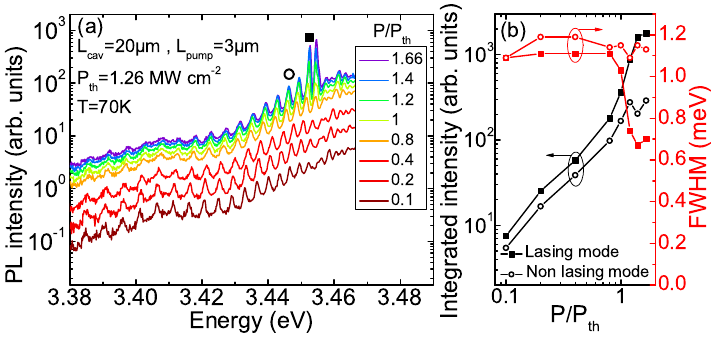}} 
  \caption{Laser operation of the $20\ \mu m$-long cavity at 70K. (a)~Power-dependent PL spectra across the lasing threshold; (b)~intensity (black) and linewidth (red) of the lasing and the non-lasing modes. }
  \label{fig:laser operation}
\end{figure}

 Figure~\ref{fig:laser operation}(a) shows a series of power-dependent PL spectra recorded at the DBR position for a $20\ \mu m$-long cavity excited by a $3\ \mu m$-long pump spot positioned at the cavity center. The spectra exhibit FP modes over a large energy range. For the mode at $3.452\ eV$ (square mark), the intensity rapidly increases by about one order of magnitude across a threshold value $P_{th}=1.26\ MW.cm^{-2}$ of the excitation power (Fig.~\ref{fig:laser operation}(b)). This mode also evidences a line narrowing beyond threshold, with a decrease of the full width at half maximum (FWHM) from $1.1 \ meV$ down to $0.65\ meV$. These two features evidence the laser operation of the device. On the contrary, for a non-lasing FP mode at negative detuning (open circle mark), the intensity undergoes a linear increase vs the pump power and the linewidth remains almost constant with a FWHM of $1.15\ meV$.

\subsection{Polariton dispersion} 

\begin{figure}
\resizebox{\hsize}{!}{\includegraphics{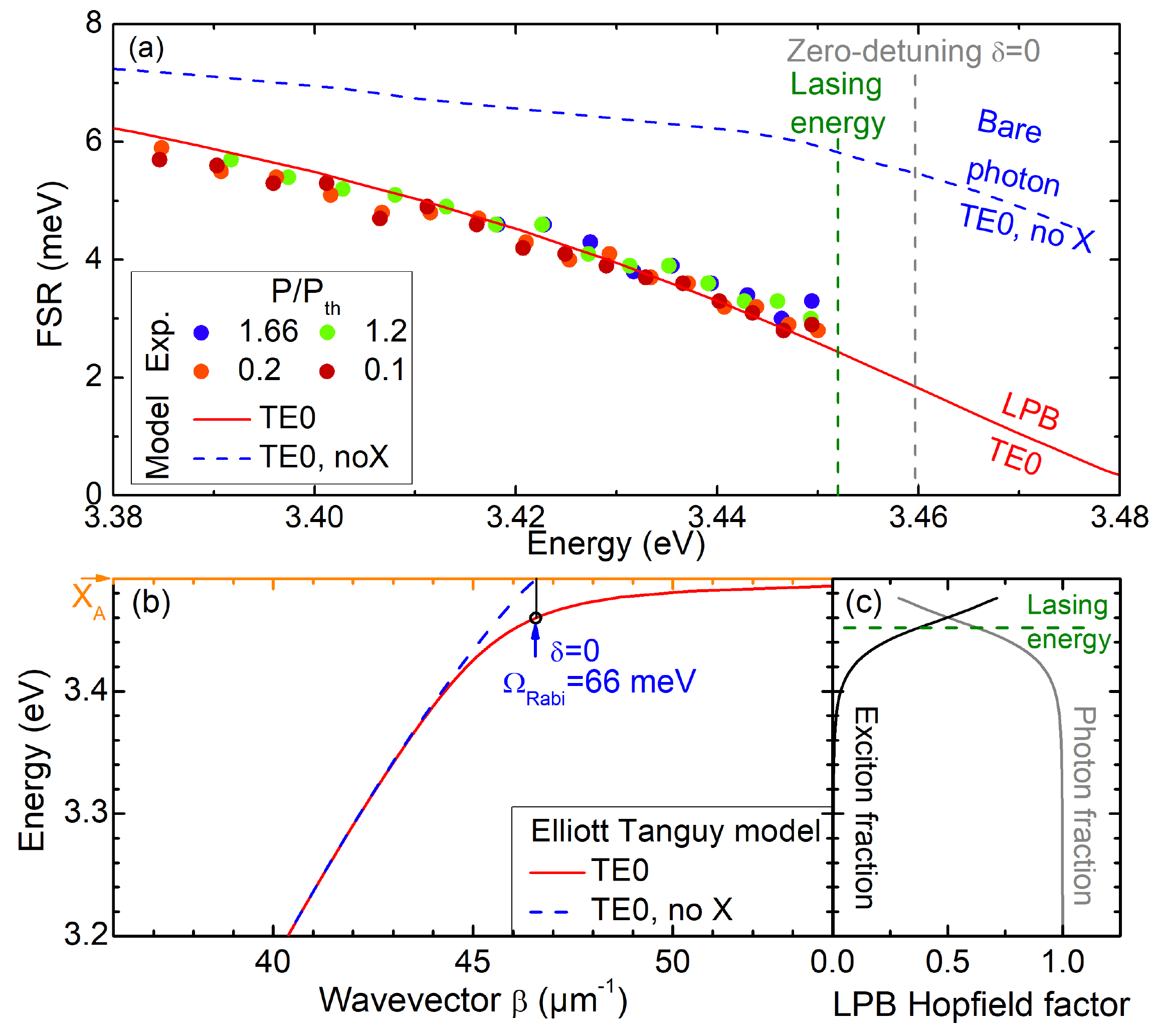}} 
  \caption{(a)~Experimental FSR vs excitation power (colored dots) and FSR calculated from LPB dispersions (plain line); (b) Corresponding polariton dispersion (Elliott-Tanguy model, plain red line), together with the corresponding bare waveguide mode (dashed blue line); (c) Hopfield coefficients of the LPB.}
  \label{fig:dispersion}
\end{figure}

In order to assess the strong coupling regime between photons and excitons, the most direct method would be to monitor the dispersion of polaritons using diffraction gratings on top of the core layer. This geometry was exploited for polariton waveguides based on GaAs \cite{walker_exciton_2013,Suarez-Forero_Electrically_2020}, GaN \cite {ciers_propagating_2017,Brimont_Strong_2020}, and ZnO \cite{jamadi_edge-emitting_2018}. It allowed measuring the lower polariton branch (LPB) dispersion and observing the anticrossing between lower and upper polariton branches (UPB). For laser cavities, in order to avoid the losses associated to such gratings, we can measure and model the cavity FSR (Fig.~\ref{fig:dispersion}(a)),  which is equal to $(L_{cav}/\pi) \ {\partial E_{LPB}}/{\partial \beta},$ where $E_{LPB}(\beta)$ is the polariton dispersion (Fig.~\ref{fig:dispersion}(b)). This method has been previously implemented in order to assess the strong coupling regime in the case of ZnO-nanowires \cite {vanmaekelbergh_zno_2011,versteegh_room-temperature_2012}, and ZnO polaritonic waveguides (Supplemental Material in Ref.~\citenum{jamadi_edge-emitting_2018}). Figure~\ref{fig:dispersion}(a) compares the FSRs extracted from the FP transmission spectra, below and above threshold to the FSRs extracted from calculated dispersions for the LPB and bare photonic mode, {\it i.e.} in the absence of strong-coupling (see Appendices~\ref{app:excitons} and \ref{app:dispersion} for details on calculations). The agreement between experimental and calculated FSR dispersions is obtained for the TE0 mode strongly coupled to excitons, with a reduction of the oscillator strength of $30\%$ below threshold ($0.1$ and $0.2\times P_{th}$), and $40\%$ above threshold ($P=1.66\times P_{th}$). This result demonstrates the operation in the strong coupling regime, with a slight diminution of the excitonic oscillator strength below threshold that could be related to a degradation of the structural and optical quality of the GaN active material during the deep-etching of the ridge. The zero-detuning condition ($\delta =0$) is realized at an LPB energy of $3.459\ eV$, and the exciton-photon coupling strength (Rabi splitting) amounts to $\Omega_{Rabi}=66\pm 10\ meV$ (Fig.~\ref{fig:dispersion}(b)) . The observed lasing mode lies close to zero-detuning, corresponding to an estimated exciton fraction of $40\%$ (Fig.~\ref{fig:dispersion}(c)).

\subsection{Laser losses and gain}
In order to compare conventional and polariton ridge lasers, we here follow a methodology analogous to the standard Hakki-Paoli analysis of the spectra \cite {hakki_gain_1975}; thanks to it, we can extract the total losses and, subsequently, model the gain of the active layer.
The energy-dependent FSR and finesse of the Fabry-Perot cavity are important to assess the strong coupling regime and the value of the total losses $\alpha_t$. 

\begin{figure}
\resizebox{\hsize}{!}{\includegraphics{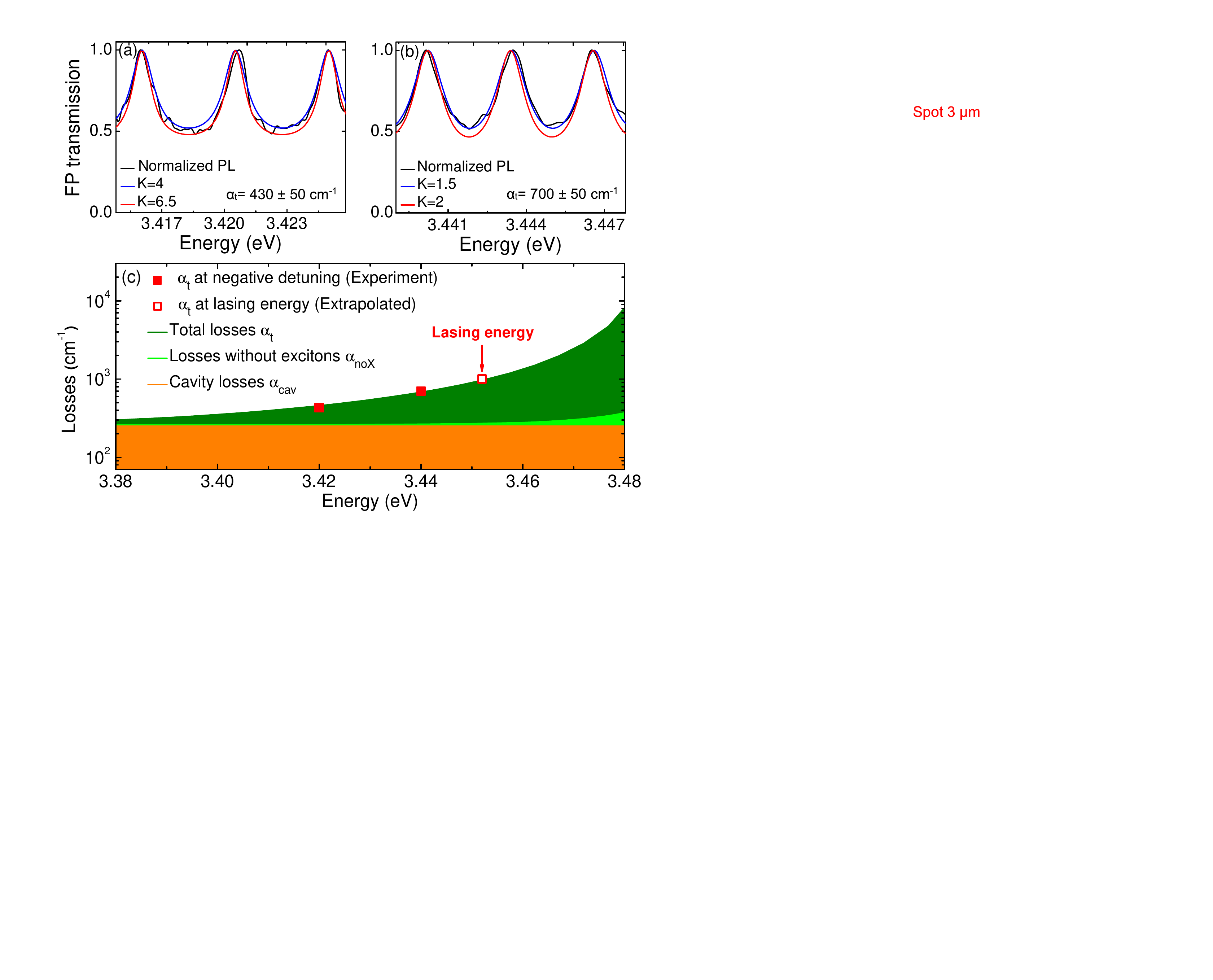}} 
  \caption{Modeling of the Fabry-Perot modes of the $20\ \mu m$-long cavity at $T=70\ K$. (a,b)~Experimental estimate of the FP transmission (black line), and its fit for various coefficient of finesse $K$ (red and blue lines); (c)~Calculated contributions to the total losses $\alpha_t$(dark green curve), from Elliott-Tanguy model, compared to experimental total losses deduced from FP transmission (red square dots).
  The open red square dot corresponds to the extrapolated total losses at the lasing energy $\alpha_t=1000\ \pm 100\ cm^{-1}$.}
  \label{fig:Study at 70K}
\end{figure}

The Fabry-Perot transmission is obtained by normalizing the PL spectrum to its envelope. Due to our collection scheme (perpendicular to the waveguide plane), the very strong spontaneous emission of the exciton reservoir under the excitation spot also contributes to the signal collected at the DBRs, even if far apart. The FP transmission can be fitted by the expression:

\begin{equation}
 T_{FP}(\lambda)=C+(1-C).\frac{1}{1+K.\sin^2(2\pi.n_g(\lambda).\frac{L_{cav}}{\lambda}+\varphi_0)}
\end{equation}

where $n_g(\lambda)=\frac{\vartheta_g}{c}$  is the group index, $\vartheta_g$ is the group velocity, $\varphi_0$ is the adjustment phase, $C$ corresponds to the background PL of the exciton reservoir and $K$ is the coefficient of finesse. The fit of the experimental spectra at two different energies at negative detuning is presented in the Figures~\ref{fig:Study at 70K}(a,b). The finesse $\mathcal{F}$ of the cavity can be deduced from the coefficient of finesse $K.$

\begin{equation}
 \mathcal{F}=\frac{\Delta \lambda }{\lambda }=\frac{\pi}{2.arcsin(\frac{1}{\sqrt{K}})}
\end{equation}

The finesse of the FP modes strongly depends on the energy.  We can deduce the total losses $\alpha_t$ from the expression of the coefficient of finesse $K$~: 

\begin{equation}
 K=\frac{4e^{-\alpha_t.L_{cav}}}{\left[1-\ e^{-\alpha_t.L_{cav}}\right]^2}  
\end{equation}

The corresponding total cavity losses $\alpha_t$ increase from $430\pm50\ cm^{-1}$ at $3.42\ eV$ to $700\pm50\ cm^{-1}$ at $3.44\ eV$ (Fig.~\ref{fig:Study at 70K}(a,b)). They can be decomposed into three contributions: 
\begin{equation}
\nonumber
\alpha_t=\alpha_{cav}+\alpha_{pol}+\alpha_{noX}.
\end{equation}
The cavity losses $\alpha_{cav}=\alpha_i+\alpha_m$ include the internal losses $\alpha_i$ (absorption by defects, roughness-induced losses) and the mirror losses $\alpha_m=-ln (R_{DBR})/{L_{cav}}$ resulting from the finite DBR reflectivity $R_{DBR}$. The losses without excitons $\alpha_{noX}$ and the polariton losses $\alpha_{pol}$ are deduced from the simulations based on the Elliott-Tanguy susceptibility (Appendix~\ref{app:dispersion}, \cite {Brimont_Strong_2020}). The three contributions are represented in Figure~\ref{fig:Study at 70K}(c). The cavity losses $\alpha_{cav}=250\ cm^{-1}$ provide a significant contribution to the total losses only at large negative detunings. We can estimate a lower bound of the DBR reflectivity $R_{DBR}=0.6$ if we neglect internal losses and therefore attribute all cavity losses to the mirrors.

Above $3.45\ eV$, the intrinsic contribution of polariton losses $\alpha_{pol}$ increases and degrades the cavity finesse (Fig.~2(a)), so that the FP transmission exhibits almost sinusoidal oscillations. This prevents a proper fit of their lineshape, so that the finesse cannot be experimentally determined in this energy range corresponding to zero and positive exciton-photon detunings. However, the good quantitative agreement between experimental and simulated total losses at negative detuning (red square dots and dark green curve in Figure~\ref{fig:Study at 70K}(c)) allows estimating the total losses at the lasing energy of about $1000\ \pm 100\ cm^{-1}$ (open red square). This value is chosen as the polaritonic gain in our model of the gain versus pump length, in the case of a polaritonic laser (Fig.~\ref{fig:Polariton laser}(b), green rectangle). When we compare the polariton laser and the conventional laser, we use the same value for the cavity losses  $\alpha_{cav}=250\ cm^{-1}$ (Fig.~\ref{fig:Polariton laser}(b), orange rectangle).

\subsection{Effect of temperature}

\begin{figure}
{\includegraphics [width=8cm]{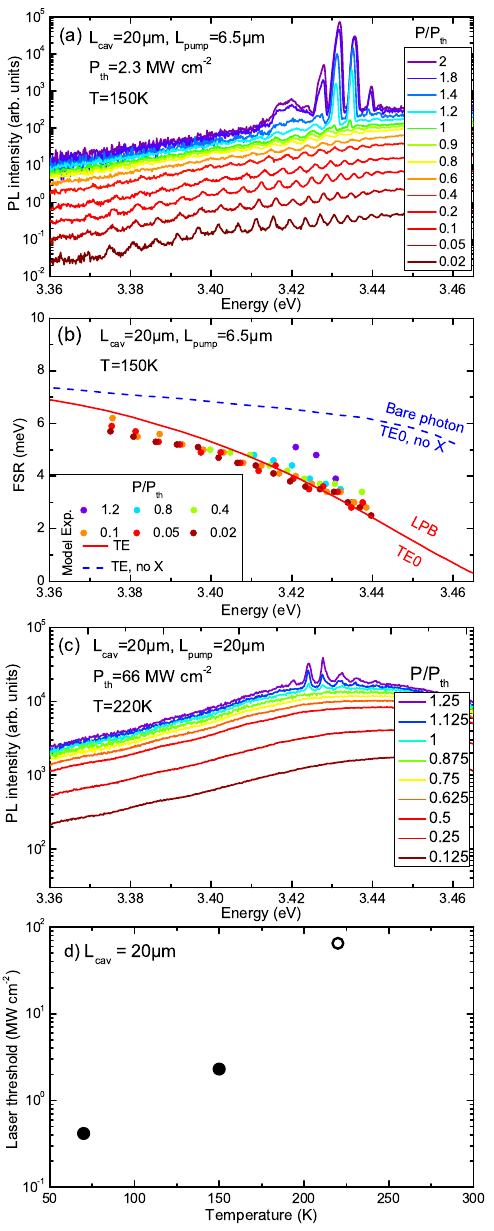}} 
  \caption{Study of the $20\ \mu m$-long cavity vs. Temperature. (a) Emission at $150\ K$ vs. pumping power. The $20\ \mu m$ length cavity is excited by a $6.5\ \mu m$ sized-pump; (b) Calculated FSR (plain lines) from dispersions for different oscillator strengths, and compared to measured Fabry-Perot modes (dots); (c) Emission at $220\ K$ vs. pumping power. The cavity $20\ \mu m$ is excited by a $20\ \mu m$ sized-pump; (d) Temperature dependence of the laser threshold for the $20\ \mu m$ length cavity. The strong coupling is verified below $T=150 \ K$ (black circles) whereas it is not assessed at $T=220 \ K$ (open black circle).
  }
  \label{fig:study vs T}
\end{figure}

The polariton laser operation has been investigated as a function of the sample temperature. The power dependent emission spectra measured at $150\ K$ on the $20\ \mu m$-long cavity with a $6.5\ \mu m$ sized-pump is shown in Figure~\ref{fig:study vs T}(a). Similarly to the $T=70\ K$ case, the Fabry-Perot modes are observed for a large energy range below threshold. The calculated FSR of the guided modes was again extracted from the dispersions based on the Elliott-Tanguy model of the dielectric susceptibility, as shown as plain lines in Figure~\ref{fig:study vs T}(b). The experimental measurements (colored dots in Figure~\ref{fig:study vs T}(b)) agree with the calculated FSRs. Above threshold a small decrease of the oscillator strength of the A, B excitons has been observed, from $-20\%$ below threshold to $-45\%$ above threshold. 

The PL spectra at $T=220\ K,$ for the same cavity excited by a $20\ \mu m$ sized-pump for a pumping power varying from $0.125$ to $1.25\times P_{th},$ are presented in Figure~\ref{fig:study vs T}(c). One can see that below threshold the Fabry-Perot modes do not appear, preventing the proper demonstration of the strong coupling at this high temperature. Still, the laser effect is maintained. The temperature increase has two notorious effects: the first impact is to shift the optimum detuning ({\it i.e.} that for which the relaxation rate is maximum) towards more and more negative values; this can be easily understood, since at higher temperatures polaritons thermalize more efficiently along the LPB and are thus able to reach more negative detunings. The second and most important effect is that the laser threshold displays a large increase with temperature, presently preventing CW lasing action up to $300\ K$, as shown in Figure~\ref{fig:study vs T}(d).

\subsection{Dependence on the pump length}

To further understand the polariton laser dynamics, and compare with conventional edge-emitting semiconductor lasers, we study the effect of the ratio of the pump length to the cavity length, in analogy with the segmented contact method for electrically injected ridge lasers \cite{Blood2003}, or the variable stripe-length method for gain measurement \cite{Shaklee1971}. The gain is fed by the exciton reservoir in a polariton laser, so that the gain length can be estimated from the spatial profile of the exciton reservoir. The measured size of the exciton reservoir is slightly larger than the length of the excitation spot, which is a signature of a $1-3\ \mu m$ spatial diffusion of the excitons independent of the excitation power in the investigated power range (Appendix~\ref{app:diffusion}).

\begin{table*}
\caption{\label{tab:table1}Losses/gain in each cavity section for a conventional edge-emitting laser and a polariton laser. $\gamma_{e-h}$ and $\gamma_{pol}$ are  the available gain from the electron-hole and from the polaritons, respectively. The cavity losses are the sum of the mirror losses $\alpha_m$ and the internal losses $\alpha_i$. $\alpha_{pol}$ and $\alpha_{noX}$ are the polariton losses and the losses without excitons, respectively.}
\begin{ruledtabular}
\begin{tabular}{cccc}

Laser&Gain section&Unpumped section&Full cavity
\\ &$L_{gain}$&$L_{cav}-L_{gain}$&$L_{cav}$\\ \hline
 Conventional laser&$+\gamma_{e-h}  \; \; (\leq \left | \alpha _0 \right | )$ & $-\alpha_0$ & $-\alpha_m-\alpha_i$\\ 
 Polariton laser&$+\gamma_{pol}-\alpha_{pol}-\alpha_{noX}$&$-\alpha_{pol}-\alpha_{noX}$ & $-\alpha_m-\alpha_i$
\end{tabular}
\end{ruledtabular}
\end{table*}

\begin{figure}
\resizebox{\hsize}{!}{\includegraphics{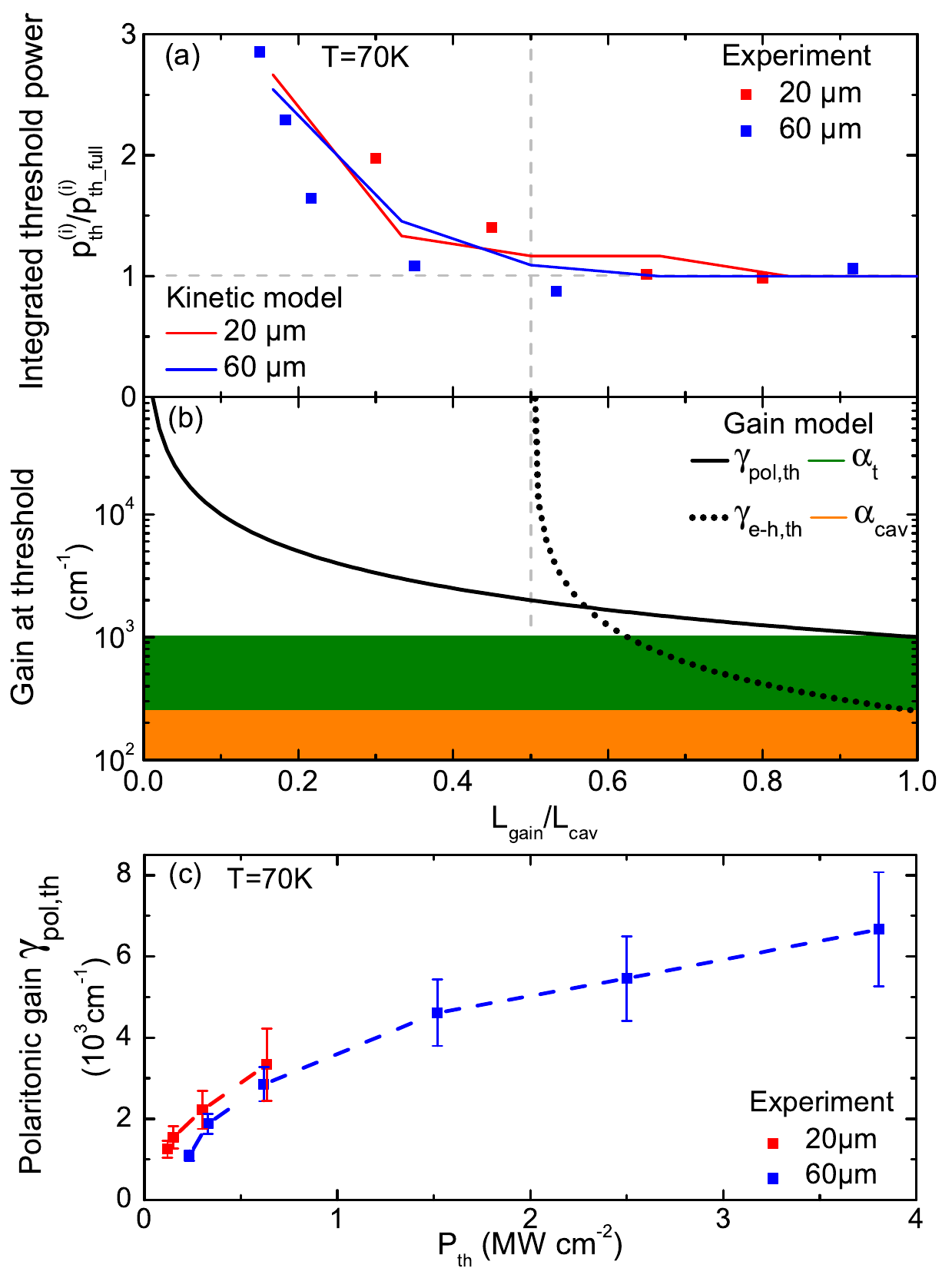}} 
  \caption{Polariton laser vs. conventional semiconductor laser. (a) Laser power at threshold $p_{th}^{(i)}$, integrated along the whole excitation length, vs gain length, normalized to its value $p_{th,full}^{(i)}$ for $L_{pump}=L_{cav}$. The $p_{th,full}^{(i)}$ are respectively $77$ and $480\ pJ/pulse$ for cavities lengths $20\ \mu m$ and $60\ \mu m$; (b) Modal gain at threshold calculated vs gain length. Plain line: case of polariton laser (Eq.~\eqref{polariton gain}). Dotted line: conventional semiconductor laser (Eq.~\eqref{conventional gain}). (c)~Polaritonic gain vs experimental power density at laser threshold $P_{th}$. Here, the dashed line is a guide for the eyes.}
  \label{fig:Polariton laser}
\end{figure}

The dependence of the spatially integrated laser power at threshold ($p_{th}^{(i)}$) versus the gain length $L_{gain}$ (Fig.~\ref{fig:Polariton laser}(a)) presents similar scaling laws for the $L_{cav}=20\ \mu m$ and $60\ \mu m$. The laser effect is achieved even for a pump length of just $10\%$ of the cavity (corresponding to a reservoir length of $15\%$ of the cavity), with only a threefold increase of the overall threshold. 

\section{Modeling and discussion}
\subsection{Model of the polariton laser kinetics}
In order to compare our experimental results with a kinetic model, we simulate the polariton relaxation in the waveguide structure with mirrors using the semiclassical Boltzmann equations for exciton-polaritons \cite{A.Kavokin2003,Solnyshkov2008} combining real and reciprocal space resolution \cite{Solnyshkov2014, Ciers2020}. The equations read:
\begin{multline}
%\begin{equation}
\frac{\partial n_{k,r}}{\partial t}=P_{k,r}-\Gamma _k n_{k,r}-n_{k,r}\sum_{k'}^{}W_{k\rightarrow k',r}(n_{k',r}+1) \\ +(n_{k,r}+1)\sum_{k'}^{}W_{k'\rightarrow k,r}n_{k',r}-\vartheta _g(k)\frac{\partial n_{k,r}}{\partial r}
\label{Boltzmann}
%\end{equation}
\end{multline}

Here, $n_{k,r}$ is the polariton (or purely exciton for large wave vectors) distribution function within a cell defined by $k$ in reciprocal space and $r$ in real space. The real space simulation, including the DBRs, is restricted to 1 dimension, whereas in reciprocal space an isotropic distribution function is accounted for (at large wave vectors, because the excitons are not quantized). The decay rate $\Gamma _k$ accounts for both the radiative and non-radiative decay rates. The decay of polaritons through the mirrors affects the spatial propagation terms at the boundaries of the system. The pumping term $P_{k,r}$ is characterized by the size of the pumping spot in real space, whereas in reciprocal space it creates a thermalized distribution of excitons within the reservoir due to the relatively rapid formation of excitons from the injected electron-hole pairs. The scattering rates $W_{k\rightarrow k',r}$ account for both exciton-exciton and exciton-phonon scattering mechanisms. They depend on the full distribution function within each spatial cell and have to be recalculated during the relaxation (for their exciton-exciton part), to account for the change of the distribution function. The term with the group velocity $\vartheta _g$ describes the polariton propagation between adjacent spatial cells. The results of this kinetic model are shown in Figure~\ref{fig:Polariton laser}.(a). A good quantitative agreement between experimental results and simulations is obtained, confirming a striking consequence of polariton lasing: the laser threshold can be reached even when pumping just one tenth of the cavity length. 

\subsection{Discussion on the polariton gain}

In a conventional ridge laser, lasing action appears due to the reciprocity between absorption and stimulated emission, and requires to achieve a population inversion in the electron and hole bands: the maximum available gain at a given energy is the product of the absorption at zero carrier density and the difference between the Fermi-Dirac distributions in the conduction and valence bands: $\gamma_{e-h}=\alpha_0 \times (f_c-f_v)$ \cite{rosencher_optolectronics_2004}; this gain is smaller than $\alpha_0$, and asymptotically equal at $T=0$. In a segmented laser with a gain section smaller than the full cavity, the gain and loss terms in both pumped and unpumped sections are compared in Table~\ref{tab:table1}, and the threshold condition writes: 
\begin{equation}
\nonumber
e^{-2{(\alpha}_m+\alpha_i).L_{cav}}.e^{-2\alpha_0.(L_{cav}-L_{gain})} .e^{2\gamma_{e-h,th}.L_{gain}}=1,
\end{equation}
where $\gamma_{e-h,th}$ is the electron-hole modal gain at threshold. Assuming $\gamma_{e-h,th}=\alpha_0$, {\it i.e.} in the $T=0$ limit, we have: 
\begin{equation}
\gamma_{e-h,th}=\frac{{(\alpha}_m+\alpha_i).L_{cav}}{{2L}_{gain}-\ L_{cav}} =\ \frac{\alpha_{cav}.L_{cav}}{2L_{gain}-\ L_{cav}} 
\label{conventional gain}
\end{equation}

Lasing action in a conventional ridge laser is obtained if more than half of the cavity is excited (Fig.~\ref{fig:Polariton laser}(b), dotted line).

In the case of a polariton laser, the polariton losses $\alpha_{pol}$ and the losses without excitons $\alpha_{noX}$ are expected in the whole cavity, {\it i.e.} even in the gain section. Moreover, the gain originates from the stimulated relaxation from the pumped exciton reservoir to the lasing polariton mode, illustrating that the polariton gain and the polariton losses are not correlated. The condition of laser oscillations at threshold writes:
\begin{multline}
\nonumber
 e^{-2(\alpha_m+\alpha_i).L_{cav}}.e^{-2(\alpha_{noX}+\alpha_{pol}).L_{cav}}.e^{2\gamma_{pol,th}.L_{gain}}=1  ,
\end{multline}
which leads to the expression of the polariton gain at threshold: 
\begin{equation}
 \gamma_{pol,th}=\frac{(\alpha_m+\alpha_i+\alpha_{noX}+\alpha_{pol}).L_{cav}}{L_{gain}}=\frac{\alpha_t.L_{cav}}{L_{gain}} 
 \label{polariton gain}
\end{equation}

The equation~\eqref{polariton gain} allows estimating the polariton gain from the previous analysis of the cavity losses. The modelled total losses $\alpha_t$ have been validated at negative detuning from the analysis of the finesse of the FP cavity, and then extrapolated to the lasing energy. At threshold, the deduced modal gain is shown on the Figure~\ref{fig:Polariton laser}(b) (plain line) as a function of the gain length. For a realistic comparison of the polariton and conventional lasing schemes, only the cavity losses are considered for the conventional ridge lasers (indicated in orange), the gain being of the same order of magnitude of the one measured for analogous nitride ridge lasers \cite {Zhang_Short_2019}. Instead, both cavity and polariton losses (indicated in green) are considered in the case of the polariton laser. Our analysis explains why a polariton laser does not require at least half of the cavity to be pumped, and provides a clear experimental evidence of this striking feature specific to polariton lasers.

Finally, the polaritonic gain at a given laser threshold is shown in the Figure~\ref{fig:Polariton laser}(c), based on the experimental data for both $L_{cav}=20\ \mu m$ and $60\ \mu m$: the two corresponding gain curves are superimposed. It is worth noting that a record modal gain of $6600\ cm^{-1}$ is reached for a short pump length. For a proper comparison with conventional ridge lasers pumped over the full cavity, it is necessary to consider the gain extrapolated for a large pump length ($1000\ cm^{-1}$). In an early GaN laser diode, a modal gain of about $400-600\ cm^{-1}$ was initially deduced from experiments \cite{Mohs1998}. More recently, in order to reach a laser operation in short cavities ($L_{cav}=45-280\ \mu m$) with high-reflectivity dielectric DBRs, a modal gain less than $100\ cm^{-1}$ has been extracted \cite{Zhang_Short_2019}. The polariton laser therefore provides a strong increase of the gain for comparable geometry and length.   

\section{conclusion}
We have studied laser cavities in GaN etched ridge structures with DBRs. A laser threshold of the order of few hundreds of $kW.cm^{-2}$ is reached, with a large gain allowing for shorter cavities compared to standard ridge GaN lasers. Beyond the demonstration of lasing, we have assessed the robustness of the strong coupling regime by comparing the cavity FSR to polariton dispersions based on the Elliott-Tanguy model \cite{Brimont_Strong_2020}. We demonstrate polariton laser operation up to $150\ K$. The laser operates on a polaritonic mode with a large exciton Hopfield factor $(40\%)$, allowing to exploit strong nonlinearities for an  intra-cavity control of the laser dynamics. Most important, we experimentally demonstrated how a ridge polariton laser is different from a standard ridge laser.  By monitoring the threshold as a function of the optical pumping length, {\it i.e.} the size of the exciton reservoir, we illustrate that a polariton laser can operate with a reservoir ({\it i.e.} a gain region) covering only $15\%$ of the full cavity, versus a theoretical lower bound of $50 \%$ for the pump length of a conventional laser. This experimental observation agrees quantitatively with simulations based on semiclassical Boltzmann equations for polaritons.

The improvement of the available gain by one order of magnitude allows for much shorter laser cavities, while the reduced pump length provides the opportunity to combine additionnal functionalities within the cavity, such as tunable absorbers or couplers for sensing or signal modulation, or topological photonic patterns, in a broad set of polaritonic materials including GaAs, ZnO, perovskites and transition metal dichalcogenides.

\section*{Acknowledgments}
The authors acknowledge fundings from the French National Research Agency (ANR-16-CE24-0021-03, ANR-11-LABX-0014, ANR-21-CE24-0019-01), the French Renatech network and the Region Occitanie (ALDOCT-001065).

\appendix
\section{Experimental setup }
\label{app:experiment}
The sample is mounted in a nitrogen-cooled cryostat with a temperature controller adjustable between $70$ and $300\ K$. The sample is excited using a pulsed laser source at $355\ nm$, emitting a pulse-width of $4\ ns$ with a repetition rate of $7\ kHz$ (Cobolt Tor). The laser power reaching the sample is controlled with a half-wave plate and a polarizer, together with optical densities. An line-shaped spot is obtained by inserting a variable slit and a cylindrical lens between the beam expander and the microscope objective. The spot width matches the ridge width. The slit is used to manually adjust the spot length $L_{pump}$. The microscope objective (20x) focuses the pump beam on the sample, and collects the emitted  signal from the cavity. The cavity is imaged onto the spectrometer’s input slit. A Charge Coupled Device (Andor Newton CCD) coupled to the $55\ cm$-long spectrometer (Horiba Jobin Yvon iHR550), with a diffractive grating with $1200$ lines per mm, allows to reconstruct the real space image.

\section{Cavity-length dependence of the FSR}
\label{app:laser60mum}

\begin{figure}[b]
\resizebox{\hsize}{!}{\includegraphics{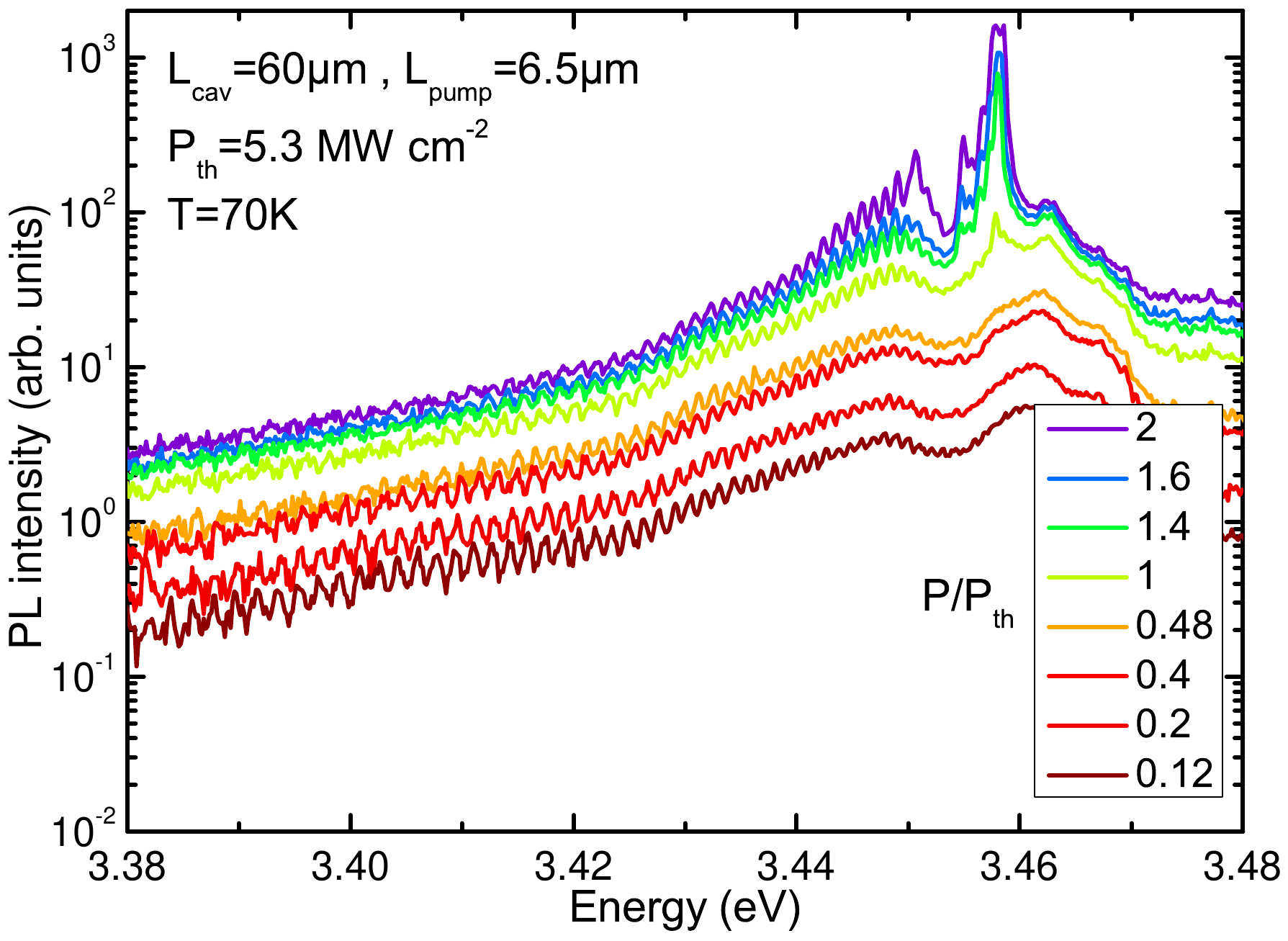}} 
  \caption{Study of the cavity $60\ \mu m$. Emission at $70\ K$ versus pumping power. The $60\ \mu m$ length cavity is excited by a $6.5\ \mu m$ sized-pump.}
  \label{fig:study of the cavity 60um}
\end{figure}

The series of PL spectra measured at $70\ K$ of a $60\ \mu m$-long cavity as a function of pumping power, for an excitation wavelength of $355\ nm$ ($L_{pump} = 6.5\ \mu m$) are presented in Figure~\ref{fig:study of the cavity 60um}. One can see also Fabry-Perot modes spreading over a large energy range, with a free spectral range (FSR) three times smaller than that of the $20\ \mu m$ long cavity. This is due to the fact that the FSR is inversely proportional to the cavity length.

\section{Exciton energies and broadenings }
\label{app:excitons}
\begin{figure}[t]
\resizebox{\hsize}{!}{\includegraphics{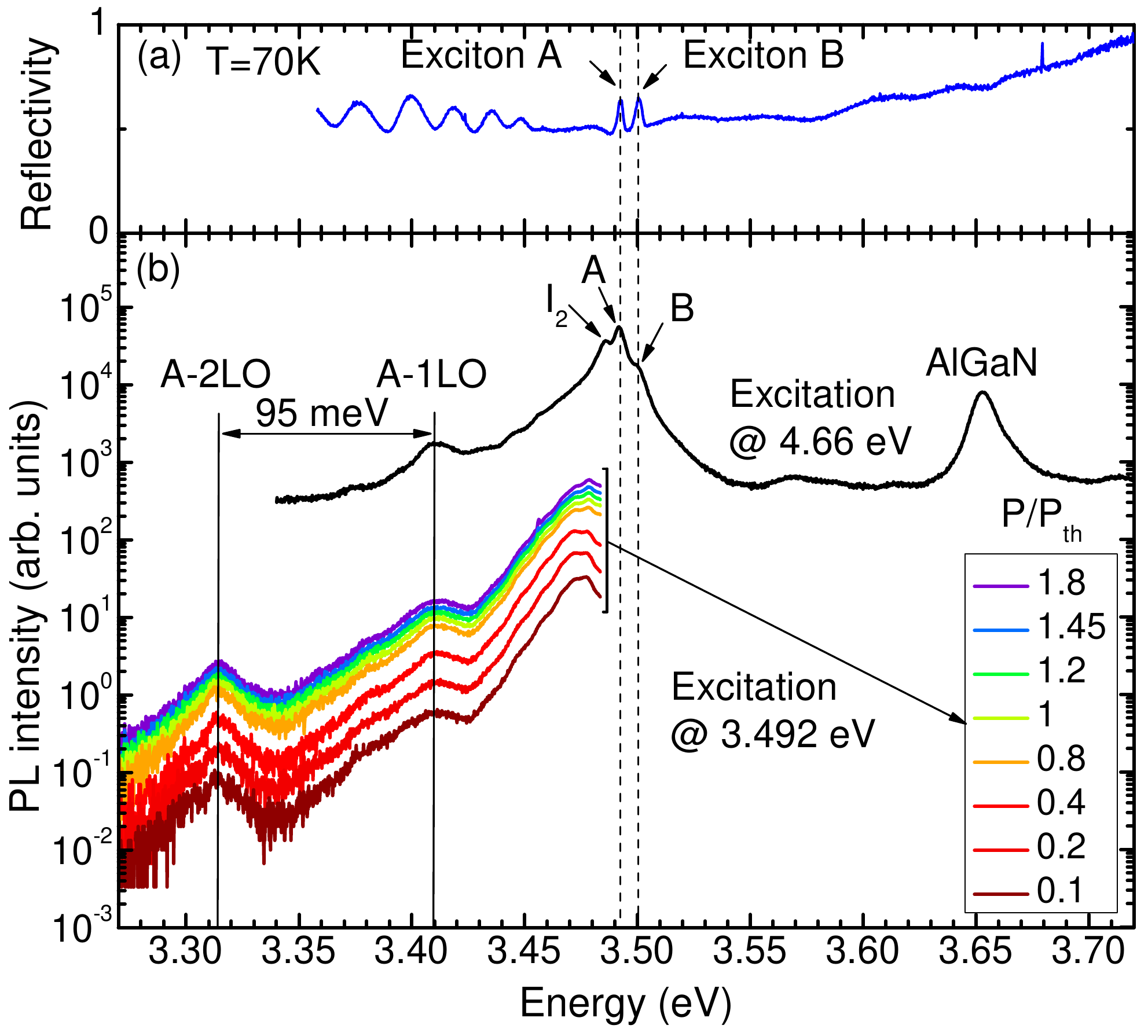}} 
  \caption{Low temperature measurements. (a) Reflectivity performed on the same sample. (b) Emission spectra at $T=70 \ K$ under non-resonant excitation at $4.66 \ eV$ ({\it i.e.} $266 \ nm$ laser, black line) and under quasi-resonant excitation at $3.492 \ eV$ vs. pump power ({\it i.e.} $355 \ nm$ laser, colored lines). The emission is measured at the position of the excitonic reservoir ($x=0\ \mu m$) of a $20\ \mu m$-long cavity.}
  \label{fig:Low T measurements}
\end{figure}

The A and B excitons energies are deduced from reflectivity experiments performed on unprocessed regions of the same sample. This leads to A and B excitons at $3.492\ eV$ and $3.5\ eV$, respectively (Fig.~\ref{fig:Low T measurements}(a)). The fit of the reflectivity provides an estimate of the homogeneous and inhomogeneous broadenings ($\gamma=1\ meV,$ $\sigma=2\ meV$). 

The PL spectra of the $20\ \mu m$-long cavity, versus pumping power, measured at $70\ K$ under the pump spot ($L_{pump}=6.5\ \mu m$, {$x=0\ \mu m$: excitonic reservoir)}, is shown in Figure~\ref{fig:Low T measurements}(b). These spectra are compared to the PL spectrum under pulsed non-resonant excitation at $266\ nm$ ($4.66\ eV$) (Fig.~\ref{fig:Low T measurements}(b), black line). We observe the A and B excitons, as well as the neutral donor bound recombination (I2 line)\cite{Leroux1999}. The peak at $3.652\ eV$ corresponds to the AlGaN cladding layer. The longitudinal-optical phonon replicas, 1LO and 2LO, of the exciton emission are observed at $3.409$ and $3.314\ eV$, respectively. Let us emphasize that the zero-phonon emission of the reservoir is at the energy of the quasi-resonant excitation laser, so that we cannot measure it in our  experiments. However the LO-phonon replicas can be easily analyzed, with constant line-shape and energy, across the whole range of excitation power. This proves that the exciton energy remains stable at all investigated powers.

\section{Modeling the polariton dispersion }
\label{app:dispersion}

\begin{figure}[t]
\resizebox{\hsize}{!}{\includegraphics{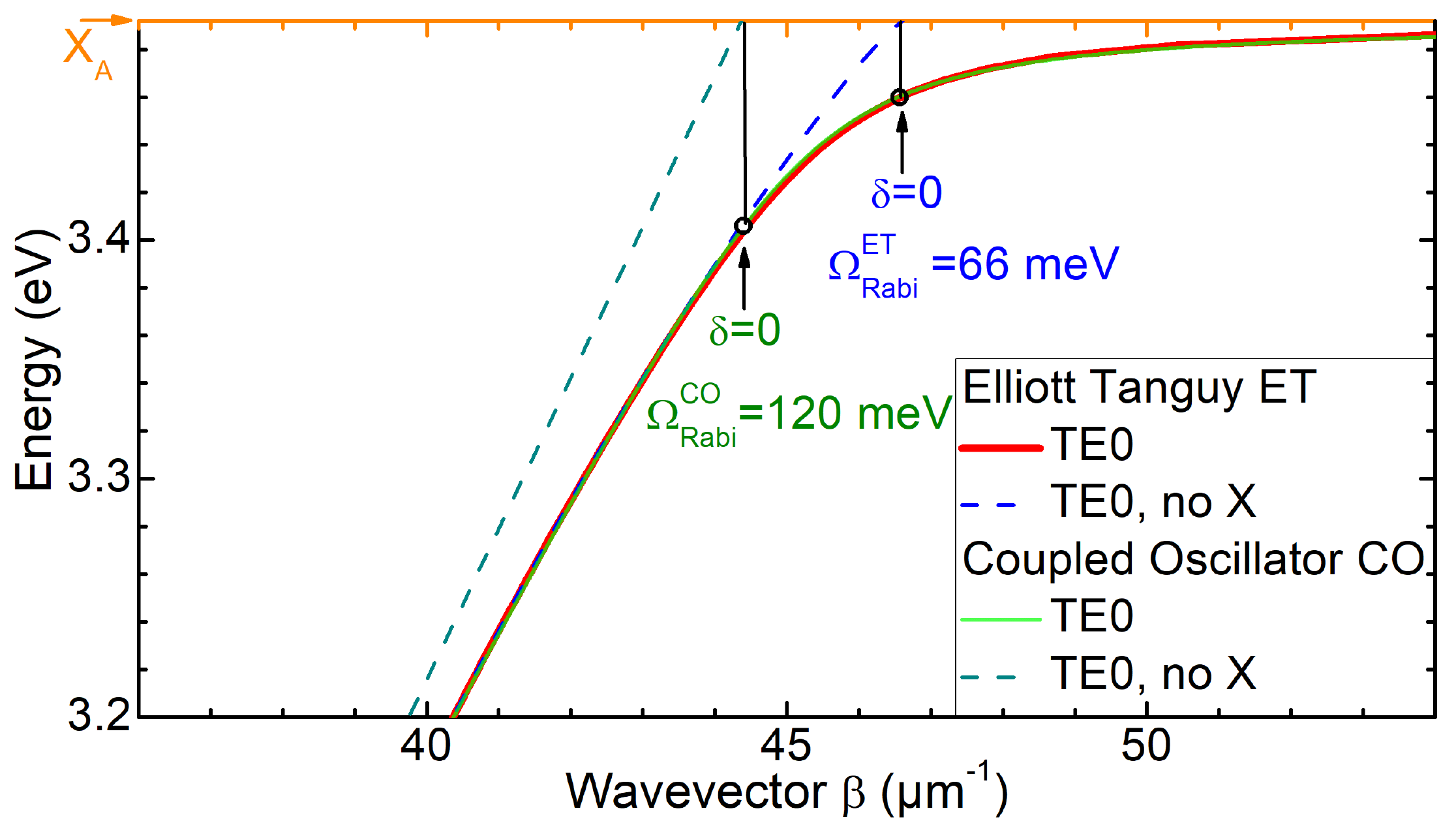}} 
  \caption{Comparison of the coupled oscillator model and the Elliott-Tanguy model of the TE0 modes of the polaritons dispersion. The corresponding bare waveguide modes are indicated as dashed lines, and the A exciton energy as a horizontal orange line. The vertical black lines indicate the zero-detuning condition for each model, and the associated Rabi splitting.}
  \label{fig:dispersion2}
\end{figure}

In order to determine the exciton-photon coupling strength, the coupled oscillator model is the simplest and most widely used, but it largely overestimates the coupling strength for waveguide polaritons. In a previous work\cite{Brimont_Strong_2020}, we proposed the so-called Elliott-Tanguy model that takes more accurately into account the dielectric susceptibility near the excitonic transition. The corresponding polariton dispersions (LPB) are presented in Figure~\ref{fig:dispersion2}, together with the dispersion of the “bare guided mode”, in the absence of exciton transitions in the dielectric function. Quantitatively, the two models lead to a zero-detuning condition ($\delta =0$) spectrally shifted with respect to each other, giving, in turn, very different estimations of the actual Rabi splitting: $120\pm10\ meV$ for the coupled oscillator model and $66\pm10\ meV$ for the Elliott-Tanguy model, corresponding to a zero-detuning condition at an LPB energy of $3.459\ eV$. All comparison of FSR dispersions with simulated dispersions are obtained within the Elliott-Tanguy model, and the material parameters are fixed as detailed in the Appendix~\ref{app:excitons}.

\section{Exciton diffusion}
\label{app:diffusion}
\begin{figure}[b]
\resizebox{\hsize}{!}{\includegraphics{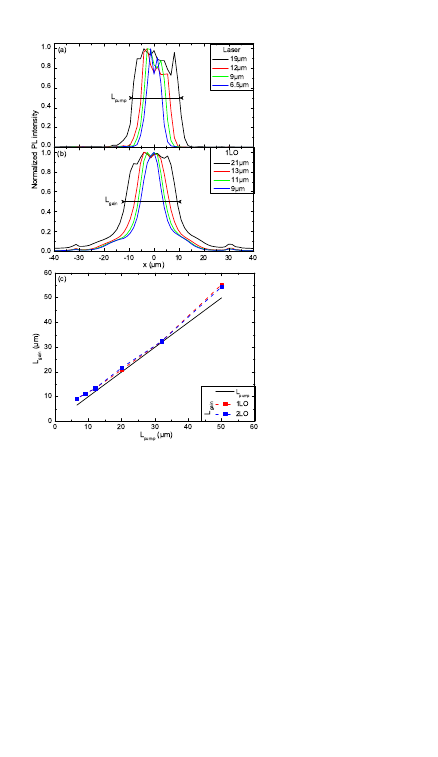}} 
  \caption{(a) The spatial profile of the laser spot for different slit opening. (b) Same as (a) but at the 1LO energy range. $L_{pump}$ and $L_{gain}$ are deduced from the FWHM. (c) Comparison of the gain length $L_{gain}$ (1LO and 2LO) and the pump length $L_{pump}$.}
  \label{fig:exciton diffusion}
\end{figure}

The effect of the pump length ($L_{pump}$) on the size of the excitonic reservoir ($L_{gain}$) is presented on the figure~\ref{fig:exciton diffusion}(a), showing the spatial profile of both the exciton reservoir and the laser at different openings of the variable slit. The FWHM allows us to estimate $L_{pump}$. In the case of polariton laser, the physical origin of the gain is the stimulated relaxation from the pumped exciton reservoir to the lasing polariton mode. Since the  energy of the pump laser is resonant with the exciton energy, we prefer estimating the gain length ($L_{gain}$) from the spatial profiles of the longitudinal-optical phonon replicas (1LO or 2LO) and their FWHM (Fig.~\ref{fig:exciton diffusion}(b)). The result of the comparison between $L_{gain}$ and $L_{pump}$ is shown in Figure~\ref{fig:exciton diffusion}(c). The exciton reservoir ($L_{gain}$) is slightly larger than the length of the excitation spot ($L_{pump}$), due to the spatial diffusion of the excitons which is of the order of $1-3\ \mu m$.

%\bibliographystyle{./apsrev4-1} 
%\bibliography{Biblio-HSouissi}

\end{document}